\documentclass[4p]{elsarticle}
\usepackage{graphicx}
\usepackage{amsmath,amssymb,bm,amsfonts}
\usepackage{slashed}

\allowdisplaybreaks[1]
\begin{document}

\begin{frontmatter}
\title{Nuclear forces and ab initio calculations\\[0.2em] of atomic nuclei}

\author[Bonn,Juelich]{Ulf-G. Mei\ss ner}

\address[Bonn]{Universit\"at Bonn,
             Helmholtz-Institut f\"ur Strahlen- und Kernphysik 
         and Bethe Center for Theoretical Physics,
             D-53115 Bonn, Germany}

\address[Juelich]{Forschungszentrum J\"ulich, 
             Institut f\"ur Kernphysik,
             Institute for Advanced Simulation, 
         and J\"ulich Center for Hadron Physics,
             D-52425 J\"ulich, Germany}
\date{\today}

\begin{abstract}
Nuclear forces and the nuclear many-body problem have been some of Gerry Brown's 
main topics in his so productive life as a theoretical physicist. In this talk, I
outline how Gerry's work laid the foundations of the modern theory of nuclear
forces and ab initio calculations of atomic nuclei. I also present some recent
developments obtained in the framework of nuclear lattice simulations.

{\rm Contribution to ``45 years of nuclear theory at Stony Brook:
              a tribute to Gerald E. Brown''}
\end{abstract}

\end{frontmatter}

\section{Prologue}

Although I have been a student of Gerry, I might be the only one who never
published a paper with him. In fact, my first project as a fresh grad student 
was to work out the $\rho$-meson coupling to the chiral bag, a hot topic
in 1982. In these days, various groups tried to develop a microscopic
theory of the so successful boson-exchange models of the nuclear forces,
based on bag, quark or Skyrme models. I became interested in this topic 
through the lectures Gerry gave at the 1981 Erice school and he offered
me to do my PhD in his group. I arrived in April 1982 at Stony Brook and 
when Gerry gave me this topic to work on, he said that we would finalize 
it in front of the fire place in the winter. Well, when I handed him a 
first draft of the paper in August, he did not even look at it in detail, 
just saying:  ``You are too fast for me, do your own stuff''. So I was pretty discouraged 
and shelved the manuscript. I thought to publish it in these proceedings, but 
unfortunately it seems to have been lost over the years. Therefore, in this 
talk I will try to give a personal recollection of Gerry's understanding of 
the nuclear force and how it influenced modern nuclear structure calculations. 
Clearly, I can only touch upon some issues here, topics like nuclear forces from 
bags or Skyrmions or the $V_{\rm low-k}$ approach are left to other speakers.   
Also, to avoid duplication, for more personal recollections I refer to my
contribution ``Chiral symmetry, nuclear forces and all that'' to the Festschrift in 
honor of his 85$^{\rm th}$ birthday,  see Ref.~\cite{Meissner:2010sa}.

\section{How to build a serious nuclear force model}

This is a condensation of the work Gerry did with many collaborators
over decades, so it is neither intended to be complete nor exhaustive.
As a matter of fact, Gerry knew all the ingredients how to model
the nuclear force \cite{geb1}. These are (I also give a few pertinent
papers  on these topics co-authored by Gerry) :
\begin{itemize}
\item[1)] {\em Chiral symmetry} \cite{geb2,geb3,geb4,geb5}:\\
Chiral symmetry fixes pion interactions to pions and matter fields
and thus relates seemingly unrelated processes, like pion-nucleon
scattering with the two-pion exchange potential and the leading three-nucleon
force of two-pion range. Gerry was one of the pioneers of exploring the
consequences of chiral symmetry in nuclear physics.
All this is now firmly rooted in the spontaneous and
explicit chiral symmetry breaking of Quantum Chromodynamics (QCD).
\item[2)] {\em Three- and four-body forces} \cite{geb2,geb3,geb6}:\\
A precise description of few-nucleon systems requires three-nucleon forces (3NFs),
see e.g. Ref.~\cite{KalantarNayestanaki:2007zi}. As examples, I mention the 
minimum in the differential cross section of low-energy neutron-deuteron
scattering or the $^3$He--$^3$H binding energy difference.  Four-nucleon forces (4NFs)
become relevant in heavier nuclei and nuclear matter, see the 
discussion in Sec.~\ref{sec:latt}.
\item[3)] {\em Two-pion exchange form pion-nucleon scattering} 
\cite{geb7,geb8,geb9,geb10}:\\
A model-independent determination of the two-pion exchange contribution to the
nuclear force is possible using dispersion relations, that allow one to connect
the processes $\pi N\to\pi N$ and $\bar NN\to\pi\pi$. This connection was
utilized in the Paris~\cite{Cottingham:1973wt} and the Stony Brook~\cite{Jackson:1975be}
potentials. The method is discussed
in great detail in one of Gerry's textbooks~\cite{Gerrybook}.
\end{itemize}
One central ingredient in a potential 
constructed along these lines is the generation of the $\sigma$--meson, that 
parameterizes the central intermediate range attraction in boson-exchange 
potentials,  through pion rescattering. For a modern look at this problem, 
I refer to Ref.~\cite{Donoghue:2006rg}. Also, there is of course a strong
overlap between these issues, as e.g. three-nucleons forces are strongly
constrained by chiral symmetry. In fact, one must admit that almost all the 
ingredients for an Effective Field Theory (EFT) approach
were already available at the time I did my graduate studies, except for
a power counting that was only formulated for multi-nucleon systems based
on chiral pion-nucleon Lagrangians by Steve Weinberg in 1990~\cite{Weinberg:1990rz}.

\section{Nuclear forces from chiral EFT}

\begin{figure}[t]
\begin{center}
\includegraphics[width=0.9\textwidth,angle=0]{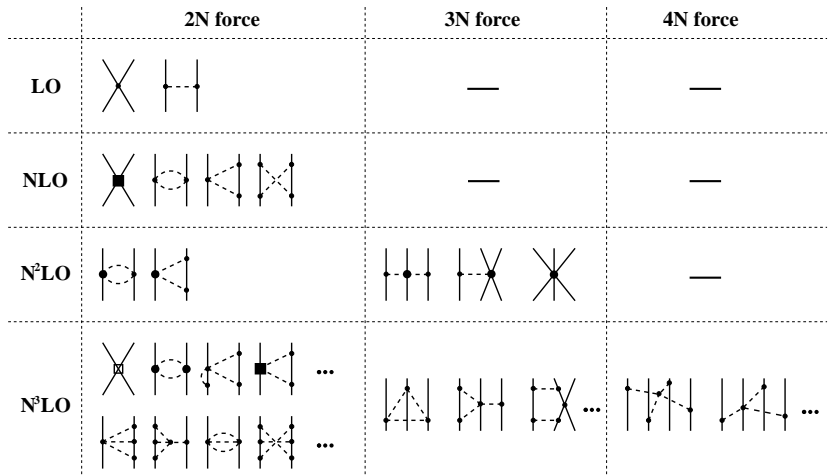}
\end{center}
\caption{\label{fig:power}
Contributions to the effective potential of the 2N, 3N and 4N forces
based on Weinberg's power counting.
Here, LO denotes leading order, NLO next-to-leading order and so on.
Dimension one, two and three pion-nucleon interactions are denoted
by small circles, big circles and filled boxes, respectively. In the
four-nucleon contact terms, the filled and open box denote two- and four-derivative
operators, respectively. Figure courtesy of Evgeny Epelbaum.}
\end{figure}

Chiral effective field theory as originally proposed by Weinberg
has become a precision tool in nuclear physics. Here, I only display
its main ingredients without detailed discussion. For an introductory
review with many references, I refer to Ref.~\cite{Epelbaum:2012vx}.
Also, I do not want to enter the issues related to the inclusion of
$\Delta$--isobars or vector mesons or some on-going attempts to improve upon
Weinberg's original work, see Ref.~\cite{Phillips:2013fia}
for a recent discussion. In the Weinberg scheme, the power counting
is applied to the few-nucleon potentials and not to the scattering
amplitudes directly.  The resulting contributions to the 2N, the 3N 
and the 4N forces are depicted in Fig.~\ref{fig:power} 
up-to-and-including-next-to-next-to-next-to-leading order (N$^3$LO). First, this 
scheme is rather predictive. While in the 2N force, one has 2,7 and
15 low-energy constants (LECs) at leading order (LO), 
next-to-leading order (NLO) and at N$^3$LO, respectively, to be determined by
a fit to data, there are only two LECs in the 3NF at N$^2$LO and none
in the 4NF. Isospin breaking through the light quark mass difference and the
electromagnetic force can also be included systematically and precisely in
this framework.  Further, note  that - consistent with 
phenomenological observations - three-nucleon forces appear
only two orders after the dominant NN forces and four-nucleon
forces are even further suppressed, appearing only at N$^3$LO.
These forces haven been successfully applied and tested in a phletora
of calculations in few-nucleon calculations and are also frequently
used (within some approximations) in many-body calculations. There are
a number of challenges remaining, like e.g. the elusive $A_y$ data in
nucleon-deuteron scattering, see e.g. Ref.~\cite{Witala:2013ioa}, or
more calculations of the electroweak response to light nuclei with
currents consistently constructed from chiral EFT, see e.g. 
Refs.~\cite{Koelling:2013rta,Girlanda:2013lra}. The interested
reader might want to consult the reviews~\cite{Epelbaum:2008ga,Machleidt:2011zz}.

\section{Quark mass dependence of the nuclear forces}

In an interesting but not much cited paper in 1987, co-authored with Herbert
M\"uther and Chris Engelbrecht, Gerry tackled the problem of how nuclear
binding depends on the number of colors, $N_C$, and the light quark masses,
$m_q$ \cite{Muther:1987sr}. For that, a one-boson-exchange model with 
varying $N_C$ and $m_q$ was
constructed. In particular, in the scalar-isoscalar central channel, the 
spectral function was reconstructed from two-pion exchange and represented
by an effective $\sigma$-meson, with mass $M_\sigma(N_C,m_q)$ and coupling
constant $g_{\sigma NN}(N_C,m_q)$. Nuclear binding was then calculated employing
Brueckner-Hartree-Fock methods. The conclusion was that ``our world is wedged into a
small corner of the two-dimensional manifold of $m_q$ versus $N_C$''. It is
interesting to re-analyze this question in chiral EFT, which allows for a
more systematic approach to possible quark mass variations. For recent
works on nuclear forces at large-$N_C$, I refer to 
Refs.~\cite{Cohen:2013tya,Phillips:2013rsa,CalleCordon:2009ps} and references therein.

\begin{figure}[t]
\includegraphics[width=5.20cm,angle=0]{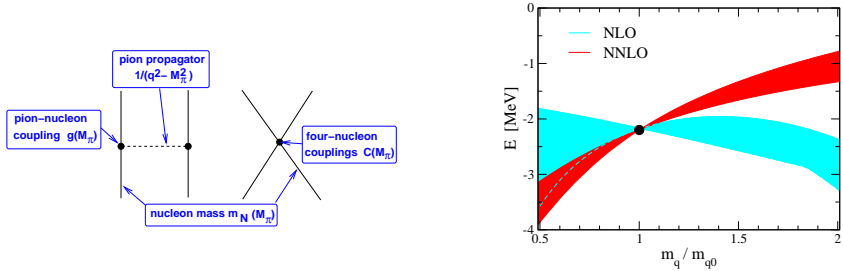}

\vspace{-3.02cm}

\hspace{6.5cm}
\includegraphics[width=4.5cm,angle=0]{BE_RunTPE.eps}
\caption{\label{fig:var}Left: Graphical representation of 
the pion (quark) mass dependence of the
nuclear force at LO (as explained in the text). Right: Quark mass dependence
of the deuteron binding energy at NLO and NNLO. Here, $m_{q0}$ denotes the
physical value of the light quark mass.}
\end{figure}

As discussed before, nuclear  forces in chiral EFT are given in terms of
pion-exchange contributions and short-distance multi-nucleon operators. The
ingredients to work out the quark mass dependence of these forces can be
easily understood from the LO potential shown in the left panel of
Fig.~\ref{fig:var}. First, there is an {\em explicit} pion mass dependence through
the pion propagator in the one- and two-pion exchanges $\sim 1/(q^2-M_\pi^2)$ 
and, second, there is a variety of {\em implicit} quark mass dependences through the 
nucleon mass, $m_N(M_\pi)$, the pion-nucleon coupling, $g(M_\pi)$,
and the LECs accompanying the multi-nucleon operators, $C(M_\pi)$. 
Note that because of the Gell-Mann--Oakes--Renner relation,
$M_\pi^2 \sim m_q$ (we work in the isospin limit $m_q=m_u=m_d$), 
one can use the notions quark and pion mass dependence synonymously.
To discuss the quark mass dependence of hadron properties, one
introduces the so-called $K$-factors, $\delta O_H/\delta m_f = K_H^f \,
(O_H/m_f)$, with $O_H$ some hadronic observable and flavor $f = u,d,s$.
The pion and nucleon properties (masses, couplings, etc.) can be 
obtained from lattice QCD data combined with chiral perturbation
theory, for details see Ref.~\cite{Berengut:2013nh}. For the contact
interactions, one has to resort to modeling, using the fairly
successful concept of resonance saturation of the LECs. This should
eventually be overcome by corresponding lattice simulations. Putting all this
together, one finds for the NN S-wave scattering lengths and the deuteron
binding energy (see the right panel of Fig.~\ref{fig:var}) \cite{Berengut:2013nh}
\begin{equation}
K_{a, 1S0}^q = 2.3^{+1.9}_{-1.8}~, \quad K_{a, 3S1}^q = 0.32^{+0.17}_{-0.18}~,
\quad K_{\rm B(deut)}^q = -0.86^{+0.45}_{-0.50}~.
\label{eq:K}
\end{equation}
A few remarks are in order. First, these $K$-factors are of natural size.
Second, the sizeable uncertainties are due to the modeling of the contact
terms by resonance saturation. Third, as Gerry told us, the nuclear forces
are very sensitive to variations in the quark mass. Interestingly, the
deuteron appears to be stronger bound as the light quark mass decreases,
consistent with the expected binding energy in the chiral limit, $E \sim
F_\pi^2/m_N \sim 10\,$MeV. Using the $K$-factors given in Eq.~(\ref{eq:K})
combined with corresponding $K$-factors for light nuclei
\cite{Bedaque:2010hr}, one can deduce limits on the possible quark mass
variations from the abundances of the elements generated in the Big Bang,
$\delta m_q/m_q = (2\pm 4)\%$ (in the isospin limit). However, as pointed
out in Ref.~\cite{Bedaque:2010hr}, the neutron lifetime, that is sensitive
to isospin-violating effects, leads to an even stronger constraint. Under
the sensible assumption that all lepton and quark masses vary with the
VEV of the Higgs field $v$, one finds $|\delta v/v| = |\delta m_q/m_q| \leq
0.9\%$ \cite{Berengut:2013nh},  which lends some credit to anthropic 
considerations in nucleosynthesis.

\section{Ab initio calculations of atomic nuclei}
\label{sec:latt}

There are two different venues to tackle
the nuclear many-body problem, that is nuclei with atomic number $A\geq 5$.
Either one utilizes the forces from EFT within a conventional, well established
many-body technique (no-core-shell-model, coupled cluster approach, etc.)
or one develops a novel scheme that combines these forces with Monte Carlo
methods that are so successfully used in lattice QCD. This novel scheme is
termed ``nuclear lattice simulations'' and has recently enjoyed 
wide recognition  as the first ever {\sl ab initio} calculation of the Hoyle state in
$^{12}$C has been performed \cite{Epelbaum:2011md}. Space is too short for a
detailed exposition of this method, so I rather make some general remarks
and discuss some very recent results obtained in this framework.

In the nuclear lattice EFT approach, 
space-time is discretized in Euclidean time on a torus of volume
$L_s\times L_s\times L_s\times L_t$, with $L_s (L_t)$ the side 
length in spatial (temporal) direction. The minimal distance
on the lattice, the so-called lattice spacing, is $a$ ($a_t$)
in space (time). This entails a maximum momentum on the lattice,
$p_{\rm max} = \pi/a$, which serves as an UV regulator of the theory.
The nucleons are point-like particles residing on the lattice sites,
whereas the nuclear interactions (pion exchanges and contact terms)
are represented as insertions on the nucleon world lines using standard
auxiliary field representations. The nuclear forces have an approximate
spin-isospin SU(4) symmetry (Wigner symmetry) \cite{Wigner:1936dx}
that is of fundamental importance in suppressing the malicious sign oscillations 
that plague any Monte Carlo (MC) simulation of strongly interacting fermion systems 
at finite density. 
\begin{figure}[t]
\begin{center}
\includegraphics[width=8cm]{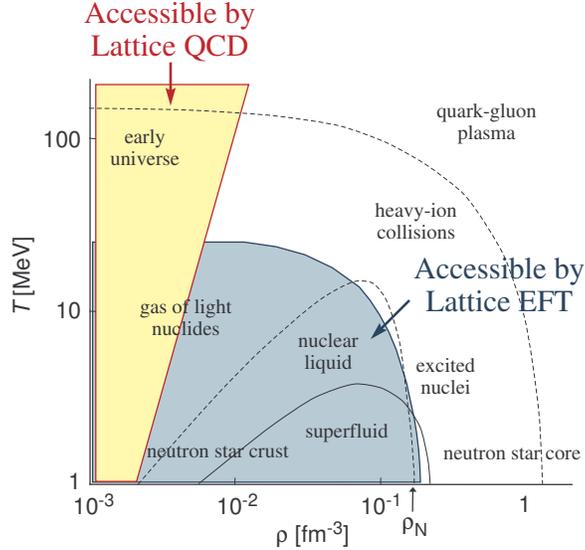}
\caption{\label{fig:phasedia}
Nuclear phase diagram as accessible by lattice QCD
and by nuclear lattice EFT. Figure courtesy of Dean Lee.
}
\vspace{-8mm}

\end{center}
\end{figure}
For this reason, nuclear lattice simulations allow access to a large part
of the phase diagram of QCD, see Fig.~\ref{fig:phasedia}, whereas 
calculations using  lattice QCD are limited to finite temperatures and
small densities (baryon chemical potential). Here, I will
concentrate on the calculation of the ground state properties and excited
states of atomic nuclei with $A \leq 28$. 

Without going into any further details, I now present some recent results concerning the 
spectrum and structure of $^{12}$C, the fate of carbon-based life as a function of the
fundamental parameters of the Standard Model, the ground-state energies of the alpha-chain 
nuclei up to $^{28}$Si and the spectrum and structure of $^{16}$O:

\smallskip

\noindent{\em Structure and rotations of the Hoyle state}~\cite{Epelbaum:2012qn}:\\
The excited state of the $^{12}$C nucleus with $J^P = 0^+$
 known as the ``Hoyle state'' constitutes one 
of the most interesting, difficult and timely challenges in nuclear physics, as 
it plays a key role in the production of carbon via fusion of three alpha particles 
in red giant stars. In Ref.~\cite{Epelbaum:2012qn}, ab initio lattice calculations
were presented which unravel the structure of the Hoyle state, along with evidence 
for a low-lying spin-2 rotational excitation. For the $^{12}$C ground state and the 
first excited spin-2 state, we find a compact triangular configuration of alpha 
clusters. For the Hoyle state and the second excited spin-2 state, we find a 
``bent-arm'' or obtuse triangular configuration of alpha clusters. The calculated 
electromagnetic transition rates between the low-lying states of $^{12}$C have also 
been obtained at LO (higher order corrections still require improved codes). 

\smallskip

\noindent{\em The fate of carbon-based life}~\cite{Epelbaum:2012iu,Epelbaum:2013wla}:\\
An ab initio calculation of the quark mass dependence of the ground state energies 
of $^4$He, $^8$Be and $^{12}$C, and of the energy of the Hoyle state in $^{12}$C have been
performed. The sensitivity of the production rate of carbon and oxygen in red giant 
stars to the fundamental constants of nature was investigated by considering the 
impact of variations in the light quark masses and the electromagnetic fine-structure 
constant on the reaction rate of the triple-alpha process. 
We find strong evidence that the physics of the triple-alpha process is driven by 
alpha clustering, and that shifts in the fundamental parameters at the $\simeq (2 - 3)\%$ 
level are unlikely to be detrimental to the development of life. Tolerance against 
much larger changes cannot be ruled out at present, given the relatively limited 
knowledge of the quark mass dependence of the two-nucleon S-wave scattering parameters,
cf. Eq.~(\ref{eq:K}). As carbon and oxygen are essential to life as we know it, 
these findings also have implications for an anthropic view of the Universe. 

\smallskip

\begin{figure}[t]
\begin{center}
\includegraphics[width=8cm]{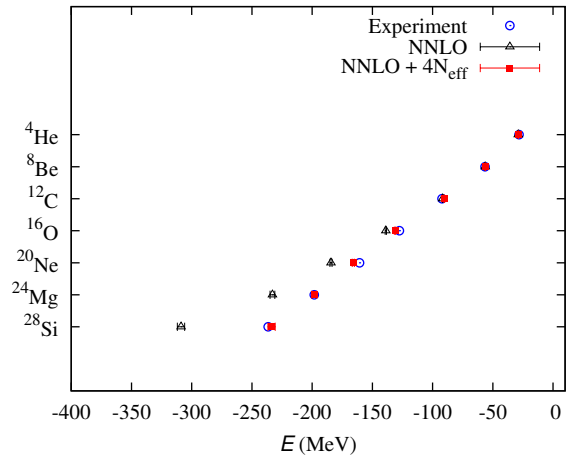}
\caption{\label{fig:alpha}
Ground-state energies of the alpha-cluster nuclei from $A=4$ to $A=28$.
The NNLO calculation is represented by the black triangles. The red squares
show the results including an effective 4N interaction, and the blue circles
are the experimental values. Figure courtesy of Dean Lee.
}
\vspace{-8mm}

\end{center}
\end{figure}
\noindent{\em Towards medium-mass nuclei}~\cite{Lahde:2013uqa}:\\
We have also extended Nuclear Lattice Effective Field Theory (NLEFT) 
to the regime  of medium-mass nuclei. To achieve that, a method which allows 
to greatly  decrease the uncertainties due to extrapolation at large Euclidean 
time was implemented. It is based on triangulation of the large Euclidean time
limit from a variety of SU(4) invariant initial interactions. 
The ground states of alpha nuclei from $^4$He to $^{28}$Si are 
calculated up to next-to-next-to-leading order in the EFT expansion.
With increasing atomic number $A$, one finds a growing overbinding as
shown in Fig.~\ref{fig:alpha}. Such effects are genuine to soft NN interactions
and also observed in other many-body calculations, 
see e.g. Refs.~\cite{Hagen:2012fb,Jurgenson:2013yya,Roth:2011ar}.
While the long-term objectives of NLEFT  are a decrease in the 
lattice spacing and the inclusion of higher-order 
contributions, it can be shown that the missing physics at NNLO can be approximated 
by an effective four-nucleon interaction. Fitting its strength to the binding
energy of $^{24}$Mg, one obtains an overall excellent description as depicted 
in Fig.~\ref{fig:alpha}.

\smallskip

\noindent{\em Spectrum and structure of $^{16}$O}~\cite{Epelbaum:2013paa}:\\
Very recently, we have performed  lattice calculations of the low-energy 
even-parity states of $^{16}$O. We find good agreement with the empirical 
energy spectrum, cf. Tab.~\ref{tab_en}, and with the electromagnetic 
properties and transition rates (after rescaling with the corrected charge
radius as detailed in \cite{Epelbaum:2013paa}). 
For the ground state, we find that the nucleons are arranged in a 
tetrahedral configuration of alpha clusters. For the first excited spin-0 state, 
we find that the predominant structure is a square configuration of 
alpha clusters, with rotational excitations that include the first spin-2 state. 
\begin{table}[h]
\begin{center}
\begin{tabular}{| c | r | r r r | r |}
\hline 
$J_n^p$ & \multicolumn{1}{c |}{LO } & \multicolumn{1}{c}{NNLO (2N)} 
& \multicolumn{1}{c}{+3N} & \multicolumn{1}{c |}{+4N$_\mathrm{eff}$} &
\multicolumn{1}{c |}{Exp} \\ \hline\hline
$0^+_1$ & $-147.3(5)$ & $-121.4(5)$ & $-138.8(5)$ & $-131.3(5)$ & $-127.62$ \\
$0^+_2$ & $-145(2)$ & $-116(2)$ & $-136(2)$ & $-123(2)$ & $-121.57$ \\
$2^+_1$ & $-145(2)$ & $-116(2)$ & $-136(2)$ & $-123(2)$ & $-120.70$ \\
\hline
\end{tabular}
\end{center}
\caption{NLEFT results and experimental (Exp) values for the lowest
  even-parity states of $^{16}$O (in MeV). 
The errors are one-standard-deviation estimates which include both statistical Monte Carlo errors and 
uncertainties due to the extrapolation $N_t^{} \to \infty$.
The combined statistical and extrapolation errors are given in
parentheses. The 
columns labeled ``LO'' and ``NNLO(2N)'' show
the energies at each order using the two-nucleon force only. The column labeled ``
+3N'' also includes the 3NF, which first appears
at NNLO. Finally, the column ``+4N$_\mathrm{eff}$'' includes the effective 
4N contribution as discussed before.  
\label{tab_en}}
\end{table}

\smallskip

\noindent I just mention some other topics under investigation within this
framework, like e.g. calculations of the equation of state of neutron matter 
and the pairing gap, a method to achieve a further reduction of the sign problem 
or setting up
methods to calculate nuclear reactions on the lattice.

\section{Some final words}

As it should have become clear, Gerry's work laid the grounds for the
modern theory of the nuclear forces and the application of these 
forces in nuclear structure calculations. In his ground-breaking paper
on the chiral EFT to the nuclear force problem, Steve Weinberg 
gives only three references, one of them being  the review Gerry
had written in 1985 with Sven-Olaf B\"ackman and Jouni 
Niskanen~\cite{Backman:1984sx} and also, he explicitly thanks Gerry
for  ``enlightening conversations on nuclear forces''. 
So clearly, Gerry has been one of the ``eagles'' of 
nuclear theory~\cite{Brown:2001xh} and his legacy will live on.

\subsection*{Acknowledgements}

I would like to thank the Dima Kharzeev, Tom Kuo, Edward Shuryak and 
Ismail Zahed for organizing this wonderful meeting. I would also like
to thank all my collaborators for sharing their insights into the topics
discussed here and Andreas Wirzba for drawing my attention to Ref.~\cite{Muther:1987sr}.
This work was supported in part by DFG, HGF, BMBF, NSFC and the EU.


\begin{thebibliography}{99}

\bibitem{Meissner:2010sa}
  U.-G.~Mei{\ss}ner,
  ``Chiral symmetry, nuclear forces and all that,''
   in  ``From Nuclei to Stars,'' Festschrift in Honor of Gerald. E Brown,
   Sabine Lee (Ed.), World Scientific (Singapore, 2011)  
   [arXiv:1011.1343 [nucl-th]].

\bibitem{KalantarNayestanaki:2007zi}
  N.~Kalantar-Nayestanaki and E.~Epelbaum,
  Nucl.\ Phys.\ News {\bf 17} (2007) 22
  [nucl-th/0703089].

\bibitem{geb1}
G.~E.~Brown, 
Comments Nucl.\ Part.\ Phys.\  {\bf 4} (1970) 140.

\bibitem{geb2}
G.~E.~Brown, A.~M.~Green and J.~W.~Gerace, 
Nucl. Phys. {\bf A118} (1968) 435.

\bibitem{geb3}
S.~Barshay and G.~E.~Brown, 
Phys. Rev. Lett. {\bf 34} (1975) 1106.

\bibitem{geb4}
G.~E.~Brown, ``Chiral Symmetry And The Nucleon Nucleon Interaction,''
in Rho M., Wilkinson D. (Eds.): Mesons In Nuclei, Vol.I*,
North-Holland Publ. Co. (Amsterdam, 1979).

\bibitem{geb5}
 J.~W.~Durso, G.~E.~Brown and M.~Saarela, 
Nucl. Phys. {\bf A430} (1984) 653.

\bibitem{geb6}
G.~E.~Brown, A.~M.~Green, J.~W.~Gerace and E.~M.~Nyman, 
Nucl. Phys. {\bf A118} (1968) 1.

\bibitem{geb7}
D.~O.~Riska and G.~E.~Brown, 
Nucl. Phys. {\bf A153} (1970) 8.

\bibitem{geb8}
G.~E.~Brown and J.~W.~Durso, 
Phys.\ Lett.\ B {\bf 35} (1971) 120.

\bibitem{geb9}
 J.~W.~Durso, M.~Saarela, G.~E.~Brown and A.~D.~Jackson, 
Nucl. Phys. {\bf A278} (1977) 445.

\bibitem{geb10}
G.~E.~Brown and R.~ Machleidt, 
Phys. Rev. {\bf C 50} (1994) 1731.


\bibitem{Cottingham:1973wt}
  W.~N.~Cottingham, M.~Lacombe, B.~Loiseau, J.~M.~Richard and R.~Vinh Mau,
  Phys.\ Rev.\ D {\bf 8} (1973) 800.

\bibitem{Jackson:1975be}
  A.~D.~Jackson, D.~O.~Riska and B.~Verwest,
  Nucl.\ Phys.\ A {\bf 249} (1975) 397.

\bibitem{Gerrybook}
G.~E.~Brown and A.~D.~Jackson, ``The nucleon-nucleon interaction,''
North-Holland Publ. Co. (Amsterdam, 1976).

\bibitem{Donoghue:2006rg}
  J.~F.~Donoghue,
  Phys.\ Lett.\ B {\bf 643} (2006) 165
  [nucl-th/0602074].

\bibitem{Weinberg:1990rz}
  S.~Weinberg,
  Phys.\ Lett.\  B {\bf 251} (1990) 288.

\bibitem{Epelbaum:2012vx}
  E.~Epelbaum and U.-G.~Mei{\ss}ner,
  Ann.\ Rev.\ Nucl.\ Part.\ Sci.\  {\bf 62} (2012) 159
  [arXiv:1201.2136 [nucl-th]].

\bibitem{Phillips:2013fia}
  D.~R.~Phillips,
  PoS CD {\bf 12} (2013) 013
  [arXiv:1302.5959 [nucl-th]].

\bibitem{Witala:2013ioa}
  H.~Witala, J.~Golak, R.~Skibinski and K.~Topolnicki,
  arXiv:1310.0198 [nucl-th].

\bibitem{Koelling:2013rta}
  S.~K\"olling,
  PoS CD {\bf 12} (2013) 096.

\bibitem{Girlanda:2013lra}
  L.~Girlanda, L.~E.~Marcucci, S.~Pastore, M.~Piarulli, R.~Schiavilla and M.~Viviani,
  PoS CD {\bf 12} (2013) 027.

\bibitem{Epelbaum:2008ga}
  E.~Epelbaum, H.~W.~Hammer and U.-G.~Mei\ss ner,
  Rev.\ Mod.\ Phys.\  {\bf 81} (2009) 1773
  [arXiv:0811.1338 [nucl-th]].

\bibitem{Machleidt:2011zz}
  R.~Machleidt and D.~R.~Entem,
  Phys.\ Rept.\  {\bf 503} (2011) 1
  [arXiv:1105.2919 [nucl-th]].

\bibitem{Muther:1987sr}
  H.~M\"uther, C.~A.~Engelbrecht and G.~E.~Brown,
  Nucl.\ Phys.\ A {\bf 462} (1987) 701.

\bibitem{Cohen:2013tya}
  T.~D.~Cohen and V.~ec.~Krejcirik,
  Phys.\  Rev.\  C 88, {\bf 054003} (2013)
   [Phys.\ Rev.\ C {\bf 88} (2013) 054003]
  [arXiv:1308.6771 [nucl-th]].


\bibitem{Phillips:2013rsa}
  D.~R.~Phillips and C.~Schat,
  Phys.\ Rev.\ C {\bf 88} (2013) 034002
  [arXiv:1307.6274 [nucl-th]].

\bibitem{CalleCordon:2009ps}
  A. Calle Cordon and E. Ruiz Arriola,
  Phys.~Rev.~C {\bf 80} (2009) 014002
  [arXiv:0904.0421 [nucl-th]].

\bibitem{Berengut:2013nh}
  J.~C.~Berengut, E.~Epelbaum, V.~V.~Flambaum, C.~Hanhart, U.-G.~Mei{\ss}ner, J.~Nebreda and J.~R.~Pelaez,
  Phys.\ Rev.\ D {\bf 87} (2013) 8,  085018
  [arXiv:1301.1738 [nucl-th]].

\bibitem{Bedaque:2010hr} 
  P.~F.~Bedaque, T.~Luu and L.~Platter,
  Phys.\ Rev.\ C {\bf 83}, 045803 (2011)
  [arXiv:1012.3840 [nucl-th]].

\bibitem{Epelbaum:2011md}
  E.~Epelbaum, H.~Krebs, D.~Lee, U.-G.~Mei\ss ner,
  Phys.\ Rev.\ Lett.\  {\bf 106 } (2011)  192501.
  [arXiv:1101.2547 [nucl-th]].

\bibitem{Wigner:1936dx}
  E.~Wigner,
  Phys.\ Rev.\  {\bf 51 } (1937)  106.

\bibitem{Epelbaum:2012qn}
  E.~Epelbaum, H.~Krebs, T.~A.~L\"ahde, D.~Lee and U.-G.~Mei{\ss}ner,
  Phys.\ Rev.\ Lett.\  {\bf 109} (2012) 252501
  [arXiv:1208.1328 [nucl-th]].

\bibitem{Epelbaum:2012iu}
  E.~Epelbaum, H.~Krebs, T.~A.~L\"ahde, D.~Lee and U.-G.~Mei{\ss}ner,
  Phys.\ Rev.\ Lett.\  {\bf 110} (2013) 11,  112502
  [arXiv:1212.4181 [nucl-th]].

\bibitem{Epelbaum:2013wla}
  E.~Epelbaum, H.~Krebs, T.~A.~L\"ahde, D.~Lee and  U.-G.~Mei{\ss}ner,
  Eur.\ Phys.\ J.\ A {\bf 49} (2013) 82
  [arXiv:1303.4856 [nucl-th]].

\bibitem{Lahde:2013uqa}
  T.~A.~L\"ahde, E.~Epelbaum, H.~Krebs, D.~Lee, U.-G.~Mei{\ss}ner and G.~Rupak,
  arXiv:1311.0477 [nucl-th].

\bibitem{Epelbaum:2013paa}
  E.~Epelbaum, H.~Krebs, T.~A.~L\"ahde, D.~Lee, U.-G.~Mei{\ss}ner and G.~Rupak,
  arXiv:1312.7703 [nucl-th].

\bibitem{Hagen:2012fb} 
  G.~Hagen, M.~Hjorth-Jensen, G.~R.~Jansen, R.~Machleidt and T.~Papenbrock,
  Phys.\ Rev.\ Lett.\  {\bf 109} (2012) 032502.
  
\bibitem{Jurgenson:2013yya} 
  E.~D.~Jurgenson, P.~Maris, R.~J.~Furnstahl, P.~Navratil, W.~E.~Ormand and J.~P.~Vary,
  Phys.\ Rev.\ C {\bf 87} (2013) 054312.

\bibitem{Roth:2011ar} 
  R.~Roth,  J.~Langhammer, A.~Calci, S.~Binder and P.~Navratil,
  Phys.\ Rev.\ Lett.\ {\bf 107} (2011) 072501.

\bibitem{Backman:1984sx}
  S.~O.~B\"ackman, G.~E.~Brown and J.~A.~Niskanen,
  Phys.\ Rept.\  {\bf 124} (1985) 1.

\bibitem{Brown:2001xh}
  G.~E.~Brown,
  Ann.\ Rev.\ Nucl.\ Part.\ Sci.\  {\bf 51} (2001) 1.

\end{thebibliography}
\end{document}